# Automated anatomy-based post-processing reduces false positives and improved interpretability of deep learning intracranial aneurysm detection

Jisoo Kim[1,2], Chu-Hsuan Lin[1,2], Alberto Ceballos-Arroyo[3], Ping Liu[2,4], Huaizu Jiang[3], Shrikanth Yadav[5], Qi Wan[6], Lei Qin[2,4], Geoffrey S Young[1,2]

[1]Dept of Radiology, Brigham and Women's Hospital, Boston, MA

[2]Harvard Medical School, Boston, MA

[3]Khoury College of Computer Sciences, Northeastern University, Boston, MA

[4]Dept of Imaging, Dana-Farber Cancer Institute, Boston, MA

[5]Dept of Imaging Science, Washing University in St. Louis, St. Louis, MO

[6]First Affiliated Hospital of Guangzhou Medical University, Guangzhou, China

Abstract

**Introduction**: Deep learning (DL) models can help detect intracranial aneurysms on CTA, but high false positive (FP) rates remain a barrier to clinical translation, despite improvement in model architectures and strategies like detection threshold tuning. We employed an automated, anatomy-based, heuristic-learning hybrid artery-vein segmentation post-processing method to further reduce FPs.

**Methods**: Two DL models, CPM-Net and a deformable 3D convolutional neural network-transformer hybrid (3D-CNN-TR), were trained with 1,186 open-source CTAs (1,373 annotated aneurysms), and evaluated with 143 held-out private CTAs (218 annotated aneurysms). Brain, artery, vein, and cavernous venous sinus (CVS) segmentation masks were applied to remove possible FPs in the DL outputs that overlapped with: (1) brain mask; (2) vein mask; (3) vein more than artery masks; (4) brain plus vein mask; (5) brain plus vein more than artery masks.

**Results**: CPM-Net yielded 139 true-positives (TP); 79 false-negative (FN); 126 FP.  3D-CNN-TR yielded 179 TP; 39 FN; 182 FP. FPs were commonly extracranial (CPM-Net 27.3%; 3D-CNN-TR 42.3%), venous (CPM-Net 56.3%; 3D-CNN-TR 29.1%), arterial (CPM-Net 11.9%; 3D-CNN-TR 53.3%), and non-vascular (CPM-Net 25.4%; 3D-CNN-TR 9.3%) structures. Method 5 performed best, reducing CPM-Net FP by 70.6% (89/126) and 3D-CNN-TR FP by 51.6% (94/182), without reducing TP, lowering the FP/case rate from 0.88 to 0.26 for CPM-NET, and from 1.27 to 0.62 for the 3D-CNN-TR.

**Conclusion**: Anatomy-based, interpretable post-processing can improve DL-based aneurysm detection model performance. More broadly, automated, domain-informed, hybrid heuristic-learning processing holds promise for improving the performance and clinical acceptance of aneurysm detection models.

Introduction

Intracranial aneurysms are abnormal protrusions from cerebral arteries caused by weakening of the vessel wall that may rupture, leading to subarachnoid hemorrhage, a condition with a 30-day mortality rate of 40-50 (1). Expert radiologists detect most aneurysms by visual inspection of 100s of separate images in each CTA and MRA. Because this is time-consuming, tiring, and open to error, deep learning (DL) models have been developed to assist clinicians with aneurysm detection. Some models can achieve sensitivities over 90% (2, 3), but the clinical translation of these models is hindered by high false positive (FP) rates that can reduce efficiency, overwhelm clinicians and prevent DL model acceptance. A recent meta-analysis of 43 studies reported FPs rates ranging from 0.13 to 31.8 per scan with a pooled FP rate of 16.5% (4), underscoring the need for robust FP reduction strategies.

Recently reported FP reduction methods include: adjusting confidence detection thresholds to reduce FPs/case, but at the cost of lowering sensitivity on MRA data (5, 6); and the use of a multi-dimensional convolutional neural network (CNN) to reduce the FP/case rate to 5.0 and 4.2 on internal and external tests, respectively, at a fixed sensitivity of 80% (7). While these methods produce important incremental improvement, they require manual adjustment or are tied to the specific model used. This and the intrinsically 'black-box' nature of model improvement limits the general applicability of these approaches and offer little insight into the nature of FPs, or how to systematically reduce them. A more systematic approach, grounded in domain-specific knowledge of aneurysm features and relevant anatomy, may help with these issues, but to date, no study has systematically analyzed the location, characteristics, and vascular anatomic relationships of FP and true positive (TP) aneurysm detections.

We propose an anatomy-based post-processing method with contemporary high-performing DL aneurysm detection models to reduce FP. Unlike previously reported solutions that are embedded within DL models, and thus inherently less generalizable and interpretable, our fully automated approach operates interpretably *after* detection. We demonstrate the use of segmented brain, artery, vein, and cavernous venous sinus (CVS) masks to identify and eliminate FP from model outputs while preserving sensitivity. By decoupling FP reduction from the DL model and grounding the FP reduction algorithm in relevant anatomic understanding, this also provides added transparency and generalizability, which will be critical for increasing acceptance of AI models in radiology.

Methods

Model

We compared two high performing DL models: 1) CPM-Net, a CNN-based 3D detector (8), 2) 3D-CNN-TR, a deformable 3D CNN-Transformer hybrid leveraging artery segmentation data as an auxiliary input (2). Each model was trained for 45 epochs using the PyTorch library. We trained both models using the AdamW optimizer with a steadily decreasing learning rate, from

0.0001 to 0.00001. Since both models were detectors, we represented aneurysms as their minimally bounding 3D boxes.  Specifically, the models were trained to output each aneurysm's location and size (height, width, depth). At inference time, we cropped a full volume into a set of contiguous sub-volumes to predict the location of potential aneurysms across the patient's CT scan, which we then stitched together for evaluation (2).

Data

This retrospective study complied with HIPAA regulations and was approved by the institutional IRB (#2004P002795). Written informed consent was waived. The training data for each model consisted of 1,186 CTAs with 1,373 annotated aneurysms available online (9).  Held-out private evaluation data were gathered through an internal medical record search tool that queried for the diagnosis of "cerebral aneurysm" in the records of 9 hospitals within our system from 2005-2016. A randomly selected subset of the CTAs and corresponding radiology reports were reviewed, and the aneurysms were annotated as bounding boxes on axial sub-millimeter source images by a radiologist with 10 years of experience (QW) on an internal research-based DICOM viewer. For additional validation, these were reviewed and edited by a subspecialty-trained neuroradiologist with 4 years of subspecialty experience (JK) on 3D Slicer ([http://www.slicer.org](http://www.slicer.org)) (10). For any discrepancies, a subspecialty-trained neuroradiologist of 25 years of subspecialty experience (GY) made the final decision on the presence and location of aneurysms. This evaluation data was also used for a previously reported paper on 3D-CNN-TR (2).

Artery and vein mask
The previously reported and publicly available intracranial artery-vein segmentation algorithm, developed by applying an nnUNet to 4D dynamic CTA data (11), was applied to all CTAs in the test data. This yielded full artery and vein segmentation inside and outside the brain (Fig 1A). Since CVS, an anatomically unique venous structure located at the base of the brain, overlaps spatially with a number of common aneurysm locations, we chose to subtract the CVS mask from the vein mask.

Cavernous venous sinus mask

The CVS was separately segmented using the Advanced Normalize Tool (ANT) registration framework. A previously reported CTA atlas including an annotated CVS region was designated as the template (11). The template CTA image was affine registered to each target CTA image using ANT, and the corresponding CVS region annotation was transformed to each target CTA image space using the same transform. The affine transformed CVS region box dimensions were expanded by 3.2 mm (8 pixels) in all directions to ensure inclusion of all surrounding anatomical structures (Fig 1B). A final CVS mask was produced, that included only overlapping voxels contained in both this expanded CVS region box and the venous segmentation mask (Fig 1C).

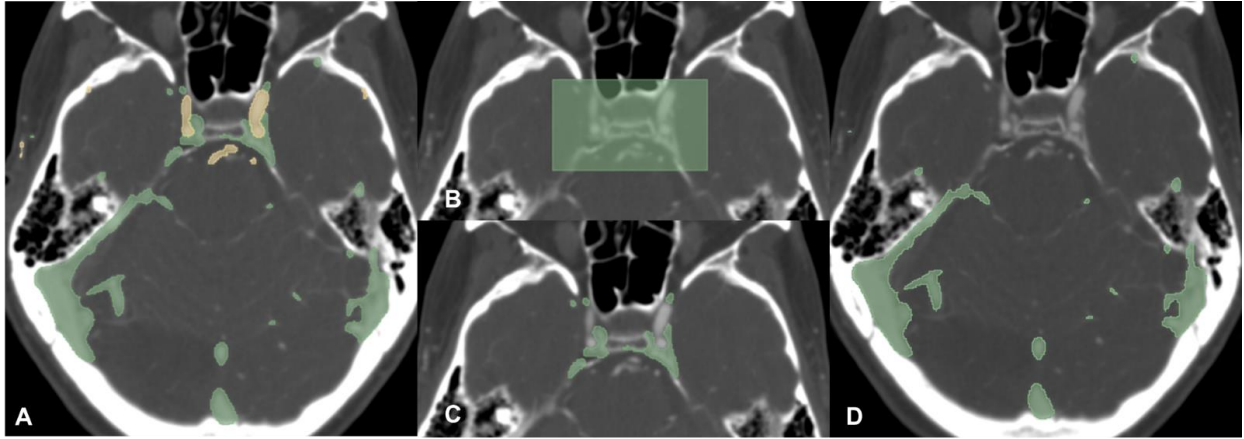

Fig 1. Arterial (yellow) and venous (green) segmentation masks of intra and extracranial structures (A). The cavernous venous sinus (CVS) region box registered on the same CTA using the ANT registration framework (B). The CVS mask defined by overlap between the CVS region box and the venous mask (C). Vein mask with CVS mask subtracted (D).

Brain mask

A brain mask was generated on all CTAs using the TotalSegmentator library (12). This mask was dilated 3.6 mm (9 pixels) in 3 dimensions to ensure inclusion of all intracranial structures. Next, the mask was combined with the CVS region bounding box to ensure inclusion of appropriate skull base structures (Fig 2).

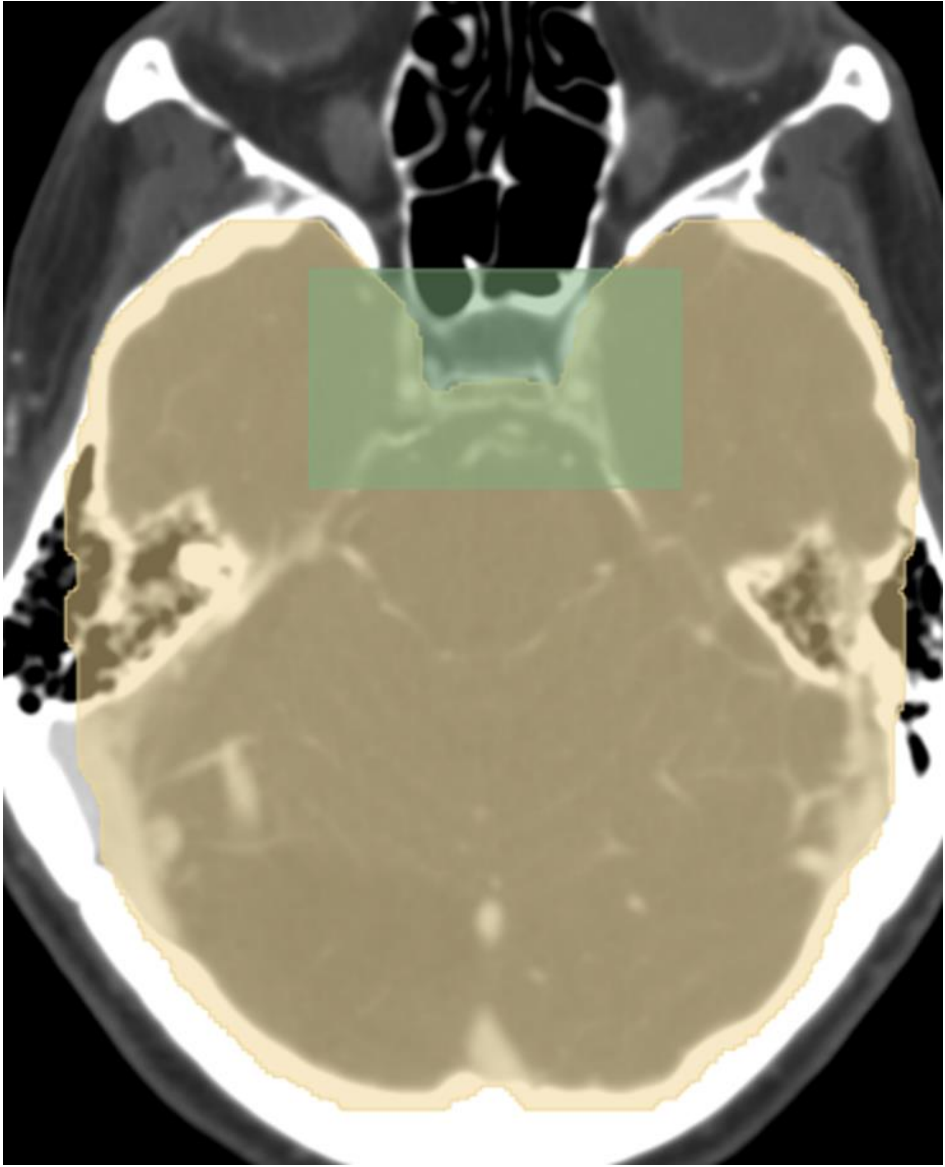

Fig 2. Brain segmentation mask using TotalSegmentator, dilated 3.6mm in each direction (yellow). The cavernous venous sinus (CVS) region bounding box from Fig 1B (green) was added to the brain mask to ensure inclusion of the appropriate skull base structures.

Post-processing

Using the artery, vein, CVS, and brain masks, five different post-processing schemes were applied to DL model outputs (which take the form of bounding boxes). Method 1 removed bounding boxes outside of the brain mask; method 2 removed bounding boxes having any overlap with the vein mask; method 3 removed bounding boxes if the bounding box overlapped with the vein mask more than the artery mask; method 4 combined methods 1 and 2; method 5 combined methods 1 and 3 (Table 1).

|  | Mask used to REMOVE bounding boxes from DL model |
|---|---|
| Method 1 | Brain |
| Method 2 | Any overlap with the vein mask |
| Method 3 | Overlap with vein mask > artery mask |
| Method 4 | Mask from Method 1 & 2 |
| Method 5 | Mask from Method 1 & 3 |

Table 1. Post-processing methods (1-5) for removing DL model output.

Analysis

A research assistant (CL) and a subspecialty-trained neuroradiologist with 5 years of subspecialty experience (PL) independently reviewed parts of the model outputs on 3D Slicer. All model outputs were subsequently reviewed by a subspecialty-trained neuroradiologist with 4 years of subspecialty experience (JK). An output was considered a TP if the bounding box was centered around the annotated aneurysm with the appropriate box size. Otherwise, output was recorded as a FP. Location and description of each FP was recorded. Outputs noted as "caliber change" included stenoses or branchpoints. Outputs noted as "vessel" included portions of small vessels that were difficult to identify as arterial or venous.

Results

Medical record search for cerebral aneurysms yielded 7749 CTAs, of which 5136 were successfully uploaded to the internal research-based DICOM viewer. A total of 143 CTAs were randomly selected. These contained 34 males and 109 females with an average age of 60.9 years. A total of 218 aneurysms were annotated, including 123 internal carotid artery (ICA), 40 middle cerebral artery (MCA), 31 anterior communicating artery (Acomm), 8 basilar, 7 superior cerebellar artery (SCA), 3 anterior cerebral artery (ACA) distal to Acomm, 2 posterior inferior cerebellar artery (PICA), and 1 posterior cerebral artery (PCA) aneurysms.

Of the 218 ground truth annotations in the test set, using a confidence threshold of 0.8, CPM-Net detected 139 TP, 79 FN, and 126 FP, for a FP rate of 0.88 FP/case. 3D-CNN-TR detected 179 TP, 39 FN, 182 FP, for a FP rate of 1.27 FP/case. Most of the FPs from CPM-Net were in venous structures (71/126, 56.3%) while those from 3D-CNN-TR were mostly in arterial structures (97/182, 53.3%) (Table 3,4). 27/99 (27.3%) CPM-Net FP and 77/182 (42.3%) of 3D-CNN-TR FP were extracranial.

The most common venous structure identified as FP were the vein of Galen (33/71, 46.4%) on CPM-Net, and extracranial veins (35/53, 66.0%) on 3D-CNN-TR. Arterial structures accounted for 15 FP on CPM-Net, and 97 FP on 3D-CNN-TR. 3D-CNN-TR identified a wider variety of normal or abnormal structures as FP, compared to CPM-Net. Structures detected as FP by 3D-CNN-TR and not by CPM-Net include basilar tip confluence (8), vertebral artery caliber change (2), PCA caliber change (1), daughter aneurysm (1), and artifacts such as aneurysm clips or calcifications (4). Structures detected as FP more commonly on 3D-CNN-TR than CPM-Net included: more cervical arteries on 3D-CNN-TR (22) than CPM-Net (2), more ICA infundibula or branchpoints on 3D-CNN-TR (37) than CPM-Net (2), more MCA branchpoints on 3D-CNN-TR (10) than CPM-Net (3), and more IAC occlusions on 3D-CNN-TR (5) than CPM-Net (1). Caliber changes at the junction of A1 and A2 segments of the ACA or other areas of ACA were identified at similar rates by 3D-CNN-TR (5) and CPM-Net (4). The two models' FP included the same number of vertebrobasilar confluences (2) and ectatic basilar arteries (1). Small vessels (intracranial and extracranial) were more commonly identified as aneurysms with 3D-CNN-TR (15) than CPM-Net (8). CPM-Net and 3D-CNN-TR both detected the choroid plexus as FPs (19 and 4). Only CPM-Net falsely identified the pineal gland (4), while only 3D-CNN-TR detected the posterior clinoid process (3).

With post-processing, use of the brain mask only (method 1) did not reduce any TP on CPM-Net or 3D-CNN-TR output. Use of the vein mask without the artery mask (method 2) reduced 10 TPs each on both CPM-Net and 3D-CNN-TR output. Use of the vein and artery masks (method 3) did not reduce any TPs. Combined post-processing (methods 4, 5) added the effects of each method.

| Post-processing | | None | Method 1 | Method 2 | Method 3 | Method 4 | Method 5 |
|---|---|---|---|---|---|---|---|
| CPM-Net | TP | 139 | 139 | 129 | 139 | 129 | 139 |
| | FP (FP/case) | 126 (0.88) | 98 (0.69) | 43 (0.30) | 48 (0.34) | 33 (0.23) | 37 (0.26) |
| | FN | 79 | 79 | 89 | 79 | 79 | 79 |
| 3D-CNN-TR | TP | 179 | 179 | 169 | 179 | 169 | 179 |
| | FP (FP/case) | 182 (1.27) | 104 (0.73) | 98 (0.69) | 116 (0.81) | 79 (0.55) | 88 (0.62) |
| | FN | 39 | 39 | 49 | 39 | 39 | 39 |

Table 2. Results of CPM-Net and 3D-CNN-TR models (confidence interval of 0.8) and post-processing methods 1-5

| | All FPs | FPs removed by |
|---|---|---|

|  |  | Method 1 | Method 2 | Method 3 | Method 4 | Method 5 |
|---|---|---|---|---|---|---|
| **Vein** | 71 | 11 | 67 | 67 | 67 | 67 |
| *Vein of Galen* | 33 | 0 | 33 | 33 | 33 | 33 |
| *Venous sinus* | 8 | 0 | 8 | 8 | 8 | 8 |
| *Cavernous venous sinus* | 3 | 0 | 0 | 0 | 0 | 0 |
| *Other intracranial veins* | 17 | 1 (foramen magnum) | 16 | 16 | 16 | 16 |
| *Extracranial veins* | 10 | 10 | 10 | 10 | 10 | 10 |
| **Artery** | 15 | 2 | 6 | 1 | 6 | 2 |
| *Cervical artery* | 2 | 2 | 2 | 1 | 2 | 2 |
| *ICA infundibulum/branch points* | 2 | 0 | 0 | 0 | 0 | 0 |
| *MCA branch points* | 3 | 0 | 1 | 0 | 1 | 0 |
| *Caliber change at A1/A2 or ACA* | 4 | 0 | 1 | 0 | 1 | 0 |
| *Vertebrobasilar confluence* | 2 | 0 | 2 | 0 | 2 | 0 |
| *ICA occlusion* | 1 | 0 | 0 | 0 | 0 | 0 |
| *Ectatic basilar artery* | 1 | 0 | 0 | 0 | 0 | 0 |
| **Vessel** | 8 | 7 | 5 | 5 | 8 | 8 |
| *Intracranial* | 1 | 0 | 1 | 1 | 1 | 1 |
| *Extracranial* | 7 | 7 | 4 | 4 | 7 | 7 |
| **Tissue** | 32 | 8 | 5 | 5 | 12 | 12 |
| *Choroid plexus* | 19 | 0 | 2 | 2 | 2 | 2 |
| *Pineal gland* | 4 | 0 | 2 | 2 | 2 | 2 |
| *Other intracranial* | 1 | 0 | 0 | 0 | 0 | 0 |
| *Extracranial* | 8 | 8 | 1 | 1 | 8 | 8 |
| **Total** | 126 | 28 | 83 | 78 | 93 | 89 |
| *Intracranial* | 99 | 1 | 66 | 62 | 66 | 62 |
| *Extracranial* | 27 | 27 | 17 | 16 | 27 | 27 |

Table 3. Analysis of false positives of the CPM-Net model output and post-processing

|  | All FPs | FPs removed by | | | | |
|---|---|---|---|---|---|---|
|  |  | Method 1 | Method 2 | Method 3 | Method 4 | Method 5 |
| **Vein** | 53 | 36 | 49 | 48 | 49 | 49 |
| *Vein of Galen* | 12 | 0 | 12 | 12 | 12 | 12 |
| *Venous sinus* | 0 | 0 | 0 | 0 | 0 | 0 |
| *Cavernous venous sinus* | 4 | 0 | 0 | 0 | 0 | 0 |
| *Other intracranial veins* | 2 | 1 (foramen magnum) | 2 | 2 | 2 | 2 |
| *Extracranial veins* | 35 | 35 | 35 | 34 | 35 | 35 |
| **Artery** | 97 | 22 | 22 | 7 | 30 | 22 |
| *Cervical artery* | 22 | 22 | 14 | 7 | 22 | 22 |
| *ICA infundibulum/branch points* | 37 | 0 | 1 | 0 | 1 | 0 |
| *MCA branch points* | 10 | 0 | 1 | 0 | 1 | 0 |
| *Caliber change at A1/A2 or ACA* | 5 | 0 | 1 | 0 | 1 | 0 |
| *Vertebrobasilar confluence* | 2 | 0 | 2 | 0 | 2 | 0 |
| *ICA occlusion* | 4 | 0 | 0 | 0 | 0 | 0 |
| *Ectatic basilar artery* | 1 | 0 | 0 | 0 | 0 | 0 |
| *Basilar tip confluence* | 8 | 0 | 1 | 0 | 1 | 0 |
| *Vertebral artery caliber change* | 2 | 0 | 0 | 0 | 0 | 0 |
| *PCA caliber change* | 1 | 0 | 1 | 0 | 1 | 0 |
| *Daughter aneurysm* | 1 | 0 | 0 | 0 | 0 | 0 |
| *Artifact (e.g. clip, calc)* | 4 | 0 | 1 | 0 | 1 | 0 |
| **Vessel** | 15 | 12 | 6 | 5 | 14 | 13 |
| *Intracranial* | 3 | 0 | 2 | 1 | 2 | 1 |
| *Extracranial* | 12 | 12 | 4 | 4 | 12 | 12 |
| **Tissue** | 17 | 8 | 7 | 6 | 10 | 10 |
| *Choroid plexus* | 4 | 0 | 2 | 2 | 2 | 2 |
| *Pineal gland* | 0 | 0 | 0 | 0 | 0 | 0 |
| *Posterior clinoid process* | 3 | 0 | 0 | 0 | 0 | 0 |
| *Other intracranial* | 2 | 0 | 0 | 0 | 0 | 0 |
| *Extracranial* | 8 | 8 | 5 | 4 | 8 | 8 |
| **Total** | 182 | 78 | 84 | 66 | 103 | 94 |
| *Intracranial* | 105 | 1 | 26 | 17 | 26 | 17 |
| *Extracranial* | 77 | 77 | 58 | 49 | 77 | 77 |

Table 4. Analysis of false positives of the 3D-CNN-TR model output and post-processing

Discussion

Reducing FPs from AI-aided intracranial aneurysm detection software outputs is critical to future successful clinical translation of CTA detection models as well as all other radiology domain AI models, because FPs waste time, lead to 'alarm fatigue', dters radiologists from using or trusting the models, and lead to unnecessary follow-up imaging and clinic visits. Improving the balance between sensitivity and FP rates may be achieved in specific instances by additional training or adjusting the model parameters, but the robustness and stability of such interventions are uncertain, and when model performance declines as a result of drift over time, it is very difficult to identify the cause. In contrast, methods based on domain-specific prior knowledge - in this case the knowledge of anatomic subdivisions embodied in the different anatomic masks - are inherently interpretable, because the anatomic accuracy of the automatically produced masks, and the structures underlying each FP can quickly be assessed by an expert. As such, hybrid post-processing approaches offer a highly interpretable method to identify the causes of drift performance of the main AI model and/or post-processing algorithm.

To design a rational hybrid method for improving DL aneurysm detection on CTA, we performed a thorough analysis of all FP cases from two of the highest-performing intracranial aneurysm detection models. Based on this FP analysis, we devised post-processing methods that were effective in markedly reducing the FP rate at minimal computational cost and no additional user time cost.

On CPM-Net and 3D-CNN-TR respectively, 27.3 and 42.3% of the FP cases occurred outside the brain, many associated with vessels such as the cervical arteries or facial veins, but also with non-vascular structures such as the lens or cartilaginous structures. This may be due to the high density of these structures on CT. All extracranial FP cases were removed with our brain mask. By adding the CVS region box to the brain mask, we were able to maintain the pipeline's sensitivity unchanged, by including TP aneurysms that are near to or within the CVS.

The most common cause of FP cases with CPM-Net and the second most common with 3D-CNN-TR were venous structures. With our venous mask which includes all venous structures except the CVS, 94.4% (67/71) and 92.5% (49/53) of the venous FPs were removed on CPM-Net and 3D-CNN-TR, respectively. We removed the CVS region from the venous mask because a significant number of TP arise in or adjacent to the CVS. Interestingly, some of the non-vascular structures were also removed with venous masks, such as the choroid plexus, which may be due to its highly vascular nature with prominent venous enhancement. Another non-vascular structure that was removed with venous masks was the pineal gland, which may be due to its proximity to prominent veins such as the internal cerebral veins.

We investigated two different methods to employ the venous mask without inadvertently removing TP aneurysm detections, in locations where the venous mask and arterial mask are often in close proximity. The first (method 2) was to remove any outputs that overlapped with the venous mask and the second (method 3) was to remove outputs that had more overlap with the venous mask than the arterial mask. The second approach effectively removed venous FPs without removing any TP.

FPs related to intracranial arterial structures were the most perplexing. With 3D-CNN-TR, this was the most common category of FP. Although some of the FPs were pathologic areas of the arteries such as occlusion, stenosis, or calcification, many of them occurred in normal-appearing branch points or infundibula. The effect of FP detection of these abnormalities is somewhat ameliorated by the fact that such abnormalities are typically of clinical significance, so drawing radiologists' attention to them (particularly to vessel occlusions), has some ancillary benefit. This is not true of normal branch points typically characterized by caliber changes (e.g., basilar tip or vertebrobasilar confluence) or mildly ectatic areas, that were also detected by both models. These were not effectively removed by our post-processing methods, although interestingly a few of them were removed with method 2 and 5, which removed output that had any overlap with the vein mask. This may in part be due to the proximity of the vein mask to the arteries in these locations.

Limitations of our study include a relatively small number of test cases. A larger scale analysis may be warranted in the future. Also, as only two models were tested, it will be intriguing to see results from other models. Other important future directions include developing hybrid models or post-processing methods that can effectively parse out non-pathologic but not perfectly cylindrical arterial structures such as branchpoints and infundibula. Additional post-processing methods can likely be developed to remove other FP cases such as the choroid plexus or pineal gland.

In conclusion, we demonstrate that use of a domain-specific anatomy-based post-processing approach markedly reduces FP rates in two high-performing contemporary DL models designed to assist radiologists in detecting intracranial aneurysms on CTA. This illustrates the potential of such hybrid heuristic-learning post-processing approaches in general. Such approaches provide the additional value of being highly interpretable by domain experts and facilitating assessment of the reasons for pipeline performance decline over time. This is an important next step toward increasing radiologist acceptance, efficiency benefit, and trust in AI assistant model, addressing critical barriers to clinical translation of radiology domain AI.

Acknowledgement

Funding for this project was provided by NIH grants R01 LM013772 and R01 LM013891.

Summary

We report application of an inherently interpretable domain-expertise informed hybrid heuristic-learning post-processing method that employs anatomic brain, artery, vein, and CVS segmentations to eliminate common FPs from the output of aneurysm detection DL models. These effectively reduced many FP cases from two of the highest-performing DL models, without reducing TP cases.

Key Points:

1. Detailed analysis of FP cases revealed their predilection for extracranial (27.3% and 42.3% on CPM-Net and 3D-CNN-TR, respectively), venous (56.3% and 29.1%), and arterial (11.9% and 53.3%) structures as well as non-vascular tissues (25.4%) with CPM-Net.

2. By understanding the model output, interpretable and effective post-processing methods were devised, reducing up to 70.6 (89/126) and 51.6% (94/182) of FP cases without reducing TP, rendering FP/case to 0.26 and 0.62, respectively on CPM-Net and 3D-CNN-TR models.

3. Interpretable anatomy-based post-processing methods can improve interpretability and confidence in model use, improve radiologist acceptance, and facilitate assessment of model drift.